\documentstyle[preprint,aps,eqsecnum,epsf]{revtex}
\tightenlines
\begin{document}
\title{Reply to Boglione and Pennington}
\author{W. Lee\cite{LANL} and D. Weingarten}
\address{IBM Research,
P.O. Box 218, Yorktown Heights, NY 10598}
\maketitle

\begin{abstract}

In a recent article we presented an argument which, we believe, shows
to be incorrect an estimate, by Boglione and Pennington, of
corrections to the valence (quenched) approximation predictions for
properties of the lightest scalar glueball.  Boglione and Pennington's
reply to our article, it appears to us, fails to address the specific
technical issues we raised.

\end{abstract}

\pacs{11.15.Ha, 12.38.Gc}

\narrowtext

In Ref.~\cite{Boglione} a formula is assumed for the errors, in valence
(quenched) approximation scalar glueball predictions, which arise from
the valence approximation's omission of glueball-quarkonium mixing.  No
derivation from QCD itself is given to suggest why this particular
formula for valence approximation errors might be expected to hold.  The
errors are then evaluated in a model.  In Ref.~\cite{Lee} we raised two
main objections to the work of Ref.~\cite{Boglione} and, as a
consequence of these objections, concluded that the claims of
Ref.~\cite{Boglione} to have shown the valence approximation to be
unreliable for properties of the lightest scalar glueball are not
correct.  Boglione and Pennington's \cite{reply} reply to our article
mentions only one of our two objections and, it appears to us, fails to
address the key technical issue on which this objection rests. To the
second problem discussed in Ref.~\cite{Lee}, Boglione and Pennington's
reply offers no response.

In the present comment we will give a brief qualitative summary of the
issues raised in Ref.~\cite{Lee}, reply to Boglione and Pennington's
\cite{reply} response to one of these issues, and, finally, consider an
additional question concerning the work of Ref.~\cite{Boglione} which
arises from the numerical lattice calculation of glueball-quarkonium
mixing in Ref.~\cite{Lee2}.

One of the objections to Ref.~\cite{Boglione} discussed in
Ref.~\cite{Lee} is a by-product of a systematic expansion derived in
Ref.~\cite{Lee} for full QCD vacuum expectation values. The first term
in this expansion is the valence approximation.  Higher terms then
give the valence approximation's error. Comparing these terms with the
formula assumed for valence approximation errors in
Ref.~\cite{Boglione}, we show in Ref.~\cite{Lee} that the error
formula of Ref.~\cite{Boglione} appears not to be a correct expression
for the error in the valence approximation as it is generally applied.

The difficulty with the error formula of Ref.~\cite{Boglione} has a
simple origin. In lattice calculations for the valence approximation
to QCD and for full QCD, with a common choice of lattice spacing and a
common choice of renormalization conditions, the valence approximation
QCD coupling constant $g_{val}$ and the full QCD coupling constant $g$
are not equal. The valence approximation $g_{val}$ may be thought of
as $g$ divided by a chromoelectric analog of a dielectric
constant taking into account the screening of $g$ by dynamical
quark-antiquark pairs present in full QCD but absent in the valence
approximation.  In the expansion for full QCD of Ref.~\cite{Lee}, the
difference between $g$ of full QCD and $g_{val}$ of the expansion's
first term is compensated by counter-terms subtracted from the
expansion's higher terms which contain closed quark loops. By choosing
$g_{val}$ entering the leading term so that full QCD and the valence
approximation obey the same renormalization conditions, the valence
approximation's accuracy is maximized and, by means of the subtracted
counter-terms, the correction terms are minimized. Phrased
differently, an optimally chosen $g_{val}$ shifts as large a
contribution as possible out of the expansion's higher order
correction terms and into the expansion's first term.

In fact, however, the expansion remains logically correct for any
choice of $g_{val}$. In particular, $g_{val}$ could be set equal to
the full QCD $g$. With this choice of $g_{val}$, the valence
approximation becomes less accurate and the expansion's correction
terms become larger. In particular, the counter-terms subtracted from
quark loops in the error terms of the expansion entirely vanish.  In
the relation assumed in Rev.~\cite{Boglione} for valence approximation
errors, no counter-terms accompany quark loops in the error terms. As
a consequence of this absence, the version of the valence
approximation to which this formula applies must have $g_{val}$ equal
to $g$.  The resulting valence approximation will be unnecessarily
inaccurate and will differ from the valence approximation as generally
applied in lattice QCD. Correspondingly, the accompanying error terms
will be unnecessarily large and will not, as claimed in
Ref.~\cite{Boglione}, be correct estimates of the errors in the
valence approximation with an optimal choice of $g_{val}$.  For
example using data of Ref.~\cite{Butler} for inverse lattice spacing
of about 1 GeV, with $g_{val}$ forced to $g$, light hadron masses and
meson decay constants differ from experiment by as much as 45\%,
rather than by less than 10\% or less than 20\%, respectively, with an
optimal choice of $g_{val}$.

Further evidence mentioned in Ref.~\cite{Lee} that the approximation
to QCD which Ref.~\cite{Boglione} identifies as the valence
approximation uses a poorly chosen $g_{val}$ is provided by the sign
predicted in Ref.~\cite{Boglione} for the error in valence
approximation mass predictions.  The valence approximation with
correctly chosen $g_{val}$, as discussed in Ref.~\cite{Weingarten},
will generally underestimate excited state masses and meson decay
constants. This expectation is supported by the data of
Refs.~\cite{Butler,Aoki}.  On the other hand, consider full QCD
expanded according to Ref.~\cite{Lee} but with the valence
approximation first term using full QCD $g$ in place of the optimal
screened $g_{val}$.  As expected for the effect of quark-antiquark
color screening and easily confirmed by an inspection of lattice data,
$g_{val}$ is less than $g$.  Thus the mass predictions of this first
term will be the valence approximation's predictions if, with no
change in the value of lattice spacing, $g_{val}$ is pushed up to
$g$. A simple agrument based on asymptotic freedom or another direct
inspection of lattice data show that raising $g_{val}$ to $g$
increases mass predictions. Thus while the valence approximation with
optimal $g_{val}$ underestimates excited state masses, the valence
approximation with $g$ in place of $g_{val}$ overestimates masses.
This second alternative is the error found in the mass predictions of
the proposed valence approximation of Ref.~\cite{Boglione}.

Boglione and Pennington \cite{reply} offer, in effect, a two part
response to the preceding discussion. First, they argue that there is a
complex relation between their expansion, relying on an effective theory
of interacting hadrons, and QCD itself, as an interaction of quarks and
gluons. As a consequence, formulating the counter-terms of
Ref.~\cite{Lee} in the language of Ref.~\cite{Boglione} and identifying
where they might occur in the arithmetic of Ref.~\cite{Boglione} is
difficult. Second, they argue that, by the nature of its construction as
an effective field theory, their theory will somewhere include the
counter-terms of Ref.~\cite{Lee} in so far as these are actually
components of QCD. In reply to the first of these comments, however, we
offer the observation of Ref.~\cite{Lee} that the counter-term to the
one-loop quark diagram is simply a derivative of the leading valence
approximation term with respect to $g_{val}$. Thus by expressing the
parameters of the effective theory as functions of $g_{val}$ it should,
in principle, be possible to express the one-loop counter-term in an
effective field theory even though quark and gluon fields themselves are
submerged.  In reply to the second comment, meanwhile, we point out that
the question is not whether the counter-terms are implicitly present but
rather exactly where they are present. With a correctly chosen $g_{val}$
the counter-terms appear subtracted from valence approximation
corrections and shift part of the would-be corrections forward into the
valence approximation. With a $g_{val}$ set equal to $g$, however, the
counter-terms are missing from the quark-loop corrections and have been
hidden, implicitly, in the valence approximation itself. The valence
approximation then becomes less accurate and the corrections become
larger.  It is this second state of affairs which we believe we have
shown occurs in the work of Ref.~\cite{Boglione}. Whatever errors
Ref.~\cite{Boglione} may find in the valence approximation as formulated
by them do not bear on the valence approximation as generally applied.

\begin{figure}
\epsfxsize=\textwidth
\epsfbox{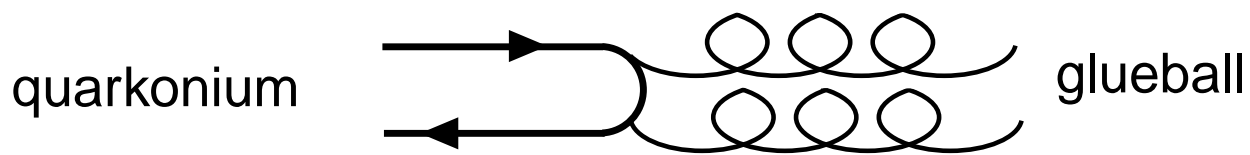}
\caption{Quarkonium-glueball mixing through quark-antiquark annihilation.}
\label{fig:mix}
\end{figure}

The other objection raised in Ref.~\cite{Lee} can be recast as the
claim that the effective field theory on which the calculation of
Ref.~\cite{Boglione} is based, according to Boglione and Pennington's
reply, is missing an important set of terms coupling glueballs and
quarkonium.  Boglione and Pennington's reply proposes that the
lightest scalar quarkonium and glueball fields entering the Lagrangian
of their effective field theory correspond to orthogonal states,
respectively a pure, ideally mixed quarkonium nonet and a pure unmixed
glueball.  Mixing among the glueball and quarkonium states then occurs
only through decay couplings linear in the scalar fields and second
order in pseudoscalar quarkonium fields.  Missing from this picture,
however, are bilinear terms in the effective Lagrangian directly
coupling the scalar glueball field to scalar quarkonium fields. Such
terms are the effective field theory's remnants of the QCD
quark-antiquark annihilation process for which a typical Feynman
diagram is shown in Figure~\ref{fig:mix}. The coefficients of these
terms can not be determined from an effective field theory and must be
taken as additional inputs from some other source.  These terms are
calculated using lattice QCD in Ref.~\cite{Lee2} and shown to make
significant contributions to quarkonium-glueball mixing. The mixing
arising from these terms explains several otherwise puzzling features
of observed scalar glueball and quarkonium data.  It is shown further
in Ref.~\cite{Lee} that mixing through these terms is quite probably
more important than mixing through pseudoscalar pairs considered in
Ref.~\cite{Boglione} if mixing through pairs is calculated according
to the systematic expansion of Ref.~\cite{Lee}. To the objection that
Ref.~\cite{Boglione} omits the dominant mixing process, Boglione and
Pennington's reply offers no response.

A third issue concerning Ref.~\cite{Boglione} arises from the
calculation in Ref.~\cite{Lee2} of the consequences of the mixing
process of Fig.~\ref{fig:mix} for the glueball decay couplings
calculated in Ref.~\cite{Sexton} and taken as input by
Ref.~\cite{Boglione}.  It is explained in Ref.~\cite{Lee2} that the
coupling calculation of Ref.~\cite{Sexton} includes the first order
effect of quarkonium-glueball mixing by the process of
Fig.~\ref{fig:mix}.  If this first order effect is removed,
Ref.~\cite{Lee2} shows, it appears possible that the remaining
coupling of the pure pseudoscalar glueball to pseudoscalar pairs may
be zero. In this case, glueball-quarkonium mixing by the mechanism
considered in Ref.~\cite{Boglione} would not only be small in
comparison to the contribution of Fig.~\ref{fig:mix}, it would be
zero.

\begin{thebibliography}{9}
\bibitem[*]{LANL} Present address:
Group T-8, Los Alamos National Laboratory, Los Alamos, NM 87545.
\bibitem{Boglione} M.\ Boglione and M.\ Pennington, Phys.\ Rev.\ Lett.\
79, 1998 (1997).
\bibitem{Lee} W.\ Lee and D.\ Weingarten, Phys.\ Rev.\ D59, 094508 (1999).
\bibitem{reply} M.\ Boglione and M.\ Pennington, submitted to Phys.\
Rev.\ D, hep-ph/9910385.
\bibitem{Lee2} W.\ Lee and D.\ Weingarten, Phys.\ Rev.\ D61, 14015 (2000).
\bibitem{Butler} F.\ Butler, H.\ Chen, J.\ Sexton, A.\ Vaccarino and 
D.\ Weingarten, Phys.\ Rev.\ Lett.\ 70 (1993) 2849; Nucl.\ Phys.\ B430, 179 (1994);
Nucl.\ Phys.\ B421, 217 (1994).
\bibitem{Weingarten} D.\ Weingarten, Nucl.\ Phys.\ B (Proc.\
Suppl.) 53, 232 (1997).
\bibitem{Aoki} S.\ Aoki, et al.\, Nucl.\ Phys.\ B (Proc.\ Suppl.) 63
(1998) 167.
\bibitem{Sexton} J.\ Sexton, A.\ Vaccarino and D.\ Weingarten,
Phys.\ Rev.\ Lett.\ 75, 4563 (1995).
\end{thebibliography}
\end{document}